# Solitary, explosive and rational solutions for nonlinear electron-acoustic waves with non-thermal electrons

S. A. El-Wakil, E. K. El-Shewy, H. M. Abd-El-Hamid, E. M. Abulwafa

*Theoretical Physics Group, Physics Department, Faculty of Science, Mansoura University, Mansoura, Egypt*

**Abstract:** A rigorous theoretical investigation has been made on electron acoustic wave propagating in unmagnetized collisionless plasma consisting of a cold electron fluid, non-thermal hot electrons and stationary ions. Based on the pseudo-potential approach, large amplitude potential structures and the existence of solitary waves are discussed. The reductive perturbation method has been employed to derive the Korteweg-de Vries (KdV) equation for small but finite amplitude electrostatic waves. An algebraic method with computerized symbolic computation, which greatly exceeds the applicability of the existing tanh, extended tanh methods in obtaining a series of exact solutions of the KdV equation. Numerical studies have been made using plasma parameters close to those values corresponding to the dayside auroral zone reveals different solutions i.e., bell-shaped solitary pulses, rational pulses and solutions with singularity at a finite points which called "blowup" solutions in addition to the propagation of an explosive pulses. The result of the present investigation may be applicable to some plasma environments, such as terrestrial magnetosphere.

## 1. Introduction

The investigation of the exact solutions of nonlinear evolution equations plays an important role in the study of nonlinear physical phenomena. Electron acoustic waves (EAWs) have been observed in the laboratory when the plasma consisted of two species of electrons with different temperatures, referred to as hot and cold electrons [1-2]. Also its propagation plays an important role not only in laboratory but also in space plasma. For example, the EAW has been used to describe the electrostatic component of the broadband electrostatic noise, in the cusp of terrestrial magnetosphere, earth bow shock, and the heliospheric termination shock, see for instances [3-5]. It has been argued that when the hot to cold electron temperature ratio is greater than 10, the electron-acoustic mode may be the principal mode of the plasma in which the restoring force comes from the pressure of the hot electrons, while the inertia comes from the mass of the cold electron component [6]. The ions play the role of a neutralizing background, i.e., the ion dynamics does not influence the EAWs because its frequency is much larger than the ion plasma frequency.

On the other hand, in practice, the hot electrons may not follow a Maxwellian distribution due to the formation of phase space holes caused by the trapping of hot electrons in a wave potential. Accordingly, in most space plasmas, the hot electrons follow the trapped/vortex-like distribution [7-11].

Recently, energetic electrons distributions are also observed in the different regions of the magnetosphere. Cairns et al [12] used non-thermal distribution of electrons to study the ion acoustic solitary structures observed by the FREJA satellite. It was shown that solitons with both positive and negative density distributions could exist. Some papers have been devoted before to study the effect of non-thermal electron distribution on the nonlinear EAWs [13-15]. Investigations of small-amplitude EAWs in collisionless plasma usually describe the evolution of the wave by nonlinear equations such as the Korteweg-de Vries (KdV), KdV-type, Zakharov--Kuznetsov (ZK), and ZK-type equations.

In the past several decades, new solutions may help to find new phenomena. A variety of powerful methods, such as inverse scattering method [16], bilinear transformation [17], tanh-sech method [18], extended tanh method [19-20], homogeneous balance method [21], Exp-function method [22] and variational iteration method [23]. Fan [24] developed a new algebraic method with computerized symbolic computation, which greatly exceeds the applicability of the existing tanh, extended tanh methods and Jacobi function expansion method in obtaining a series of exact solutions of nonlinear differential equations.

The major topic of this work is to study the existence of the electrostatic arbitrary and small amplitude solitary and other type waves in unmagnetized collisionless plasma consisting of a cold electron fluid and non-thermal hot electrons obeying a non-thermal distribution, and stationary ions. Our choice of non-thermal distribution of electrons is prompted by its convenience rather than as precise fitting of the observations. We expect that the inclusion of the non-thermal electrons will change the properties as well as the regime of existence of solitons.

This paper is organized as follows: in section 2, we present the basic set of fluid equations governing our plasma model. In section 3, the pseudo-potential was discussed. In section 4, the nonlinear small-amplitude EAWs are investigated through the derivation of a Korteweg—de Vries (KdV) equation. In Section 5, an algorithm describing the computerized symbolic computation method is presented. In Section 6, the proposed method is applied to the KdV equation. Section 7 contains the results and discussion. Finally, some conclusions are given in section 8.

## 2. Basic Equations

We consider a homogeneous system of unmagnetized collisionless plasma consisting of cold electrons fluid, non-thermal hot electrons obeying a non-thermal distribution and stationary ions. Such system is governed by the following normalized equations in one dimension [14]:

$$\frac{\partial}{\partial t} n_c(x,t) + \frac{\partial}{\partial x}[n_c(x,t) u_c(x,t)] = 0, \tag{1a}$$

$$\frac{\partial}{\partial t} u_c(x,t) + u_c(x,t) \frac{\partial}{\partial x} u_c(x,t) - \alpha \frac{\partial}{\partial x} \phi(x,t) = 0, \tag{1b}$$

with Poisson's equation

$$\frac{\partial^2}{\partial x^2}\phi(x,t)-\frac{1}{\alpha}n_c(x,t)-n_h(x,t)+(1+\frac{1}{\alpha})=0. \tag{1c}$$

The non-thermal hot electrons density $n_h(x,t)$ is given by:

$$n_h(x,t)=[1-\beta\phi(x,t)+\beta\phi(x,t)^2]\exp[\phi(x,t)], \qquad \beta=\frac{4\delta}{1+3\delta}. \tag{2}$$

In the earlier equations, $n_c(x,t)$ [$n_h(x,t)$] is the cold (non-thermal hot) electrons density normalized by equilibrium value $n_{c0}$ ($n_{h0}$), $u_c(x,t)$ is the cold electrons fluid velocity normalized by hot electron acoustic speed $C_e=\sqrt{\frac{k_B T_h}{\alpha m_e}}$, $\phi(x,t)$ is the electric potential normalized by $\frac{k_B T_h}{e}$, $\alpha=\frac{n_{h0}}{n_{c0}}$, $m_e$ is the mass of electron, $\delta$ is a parameter which determines the population of energetic non-thermal hot electrons, $e$ is the electron charge, $x$ is the space co-ordinate and $t$ is the time variable, the distance is normalized to the hot electron Debye length $\lambda_{Dh}$, the time normalized by the inverse of the cold electron plasma frequency $\omega_{ce}^{-1}$ and $k_B$ is the Boltzmann's constant. Equations (1a) and (1b) represent the inertia of cold electron and (1c) is the Poisson's equation need to make the self consistent. The hot electrons are described by non-thermal distribution given by (2).

## 3. Nonlinear Arbitrary Amplitude

To investigate the nonlinear properties of the electrostatic waves described in section 2, we must consider the nonlinear terms in (1) and (2). Therefore, in order to study the fully nonlinear (arbitrary amplitude solitary) waves, we employ here the pseudo-potential approach [25] by assuming that all dependent variables depend on a single variable $\zeta=x-Mt$, where $M$ is the Mach number (solitary wave speed normalized by hot electron acoustic speed $C_e$). Using this transformation ($\zeta=x-Mt$) along with the steady state condition ($\frac{\partial}{\partial t}=0$) and appropriate boundary conditions for localized perturbations (viz. $|\zeta|\to\infty$, $n_c=n_h=1$, $u=0$, $\phi=0$), we can reduce (1)-(2) to

$$\frac{1}{2}[\frac{d\phi}{d\zeta}]^2+V(\phi)=0, \tag{3}$$

where the potential $V(\phi)$ is given by

$$V(\phi)=(\frac{1}{\alpha}+1)\phi-(\beta\phi^2-3\beta\phi+3\beta+1)e^{\phi}+3\beta+1-\frac{M}{\alpha^2}[\sqrt{(M^2+2\alpha\phi)}-\sqrt{M^2}]. \tag{4}$$

The nonlinear equation (3) can be regarded as "energy integral" of an oscillating particle of unit mass, with velocity $\frac{d\phi}{d\zeta}$ and position $\phi$. This equation is valid for arbitrary amplitude electron-acoustic waves in the steady state. A necessary conditions for the existence of the solitary waves is that, $V(\phi=0)=0$, $\left.\frac{\partial V}{\partial \phi}\right|_{\phi=0}=0$, $\frac{\partial^2 V}{\partial \phi^2}<0$ and $V(\phi)<0$ for $0\leq\phi\leq\phi_m$.

## 4. Nonlinear Small Amplitude

According to the general method of reductive perturbation theory, we introduce the slow stretched co-ordinates [26]:

$$\tau=\varepsilon^{3/2}t, \text{ and } \xi=\varepsilon^{1/2}(x-\lambda t), \tag{5}$$

where $\varepsilon$ is a small dimensionless expansion parameter and $\lambda$ is the wave speed normalized by $C_e$. All physical quantities appearing in (1) are expanded as power series in $\varepsilon$ about their equilibrium values as:

$$n_c = 1 + \varepsilon n_1 + \varepsilon^2 n_2 + \varepsilon^3 n_3 + ..., \tag{6a}$$

$$u_c = \varepsilon u_1 + \varepsilon^2 u_2 + \varepsilon^3 u_3 + ..., \tag{6b}$$

$$\phi = \varepsilon \phi_1 + \varepsilon^2 \phi_2 + \varepsilon^3 \phi_3 + .... \tag{6c}$$

We impose the boundary conditions as $\xi\to\infty$, $n_c=n_h=1$, $u=0$, $\phi=0$.

Substituting (5) and (6) into (1) and equating coefficients of like powers of $\varepsilon$, the lowest-order equations in $\varepsilon$ lead to the following results:

$$n_1(x,t)=-\frac{\alpha}{\lambda^2}\phi_1(x,t), \text{ and } u_1(x,t)=-\frac{\alpha}{\lambda}\phi_1(x,t). \tag{7}$$

Poisson's equation gives the linear dispersion relation

$$\lambda^2 = \frac{1}{1-\beta} = \frac{1+3\delta}{1-\delta}. \tag{8}$$

Considering the coefficients of $O(\varepsilon^2)$ and eliminating the second order perturbed quantities $n_2$, $u_2$ and $\phi_2$ lead to the following KdV equation for the first-order perturbed potential:

$$\frac{\partial}{\partial \tau}\phi_1(\xi,\tau)+A\phi_1(\xi,\tau)\frac{\partial}{\partial \xi}\phi_1(\xi,\tau)+B\frac{\partial^3}{\partial \xi^3}\phi_1(\xi,\tau)=0, \tag{9}$$

where

$$A=-\frac{(3\alpha+\lambda^4)}{2\lambda} \text{ and } B=\frac{\lambda^3}{2}. \tag{10}$$

The resultant KdV equation (9) can be solved to find its solution $\phi_1(\xi,\tau)$ using different computerized symbolic computation methods [16-24].

## 5. Computerized Symbolic Computation Method

Fan [24] developed a new algebraic method with computerized symbolic computation, which greatly exceeds the applicability of the existing tanh, extended tanh methods and Jacobi function expansion method in obtaining a series of exact solutions of nonlinear differential equations.

A nonlinear partial differential equation in $\phi_1(\xi,\tau)$ has the form

$$H(\phi_1, \frac{\partial \phi_1}{\partial \tau}, \frac{\partial \phi_1}{\partial \xi}, \frac{\partial^2 \phi_1}{\partial \xi^2}, ...) = 0. \tag{11}$$

In a traveling frame of reference $\phi_1(\xi,\tau) = \Phi(\eta) = \Phi(\xi - v\tau)$ the partial differential equation may be transformed into an ordinary differential equation as

$$H(\Phi, \frac{d\Phi}{d\eta}, \frac{d^2\Phi}{d\eta^2}, ...) = 0. \tag{12}$$

Fan's computerized symbolic computational method can be used to solve (11) as follows:

**Step 1:** Transform the partial differential equation (11) into the ordinary differential equation (12) by considering the wave transformation $\phi_1(\xi,\tau) = \Phi(\eta)$, $\eta = \xi + v\tau$.

**Step 2:** Expand the solution of (12) in the form

$$\Phi(\eta) = \sum_{i=0}^{n} a_i \, \varphi(\eta)^i, \tag{13a}$$

where the new function $\varphi(\eta)$ is a solution of the following ordinary differential equation

$$\frac{d\varphi}{d\eta} = \pm \sqrt{\sum_{j=0}^{k} c_j \varphi(\eta)^j}. \tag{13b}$$

**Step 3:** Determine the parameters $n$ and $k$ by balancing the highest derivative term with the highest nonlinear term. Then obtain a relation for $n$ and $k$, from which the different possible values of $n$ and $k$ can be obtained. These values lead to the different series expansions of the solutions.

**Step 4:** Substituting (13a) and (13b) into (12) and setting the coefficients of all powers of $\varphi(\eta)^i$ and $\varphi(\eta)^i \frac{d\varphi}{d\eta}$ equal to zero will give a system of algebraic equations, from which the parameters $v$, $a_i$ ($i = 0,1,...,n$) and $c_j$ ($j = 0,1,...,k$) can be determined explicitly.

**Step 5:** Substituting the parameters $c_j$ ($j = 0,1,...,k$) obtained in step 4 into (13b), we can then obtain all the possible solutions. We remark here that the solutions of (12) depend on the explicit solvability of (13b). The solution of the system of algebraic equations will be getting a series of fundamental solutions such as polynomial, exponential, soliton, rational, triangular periodic solutions.

## 6. Explicit exact solutions for the KdV equation

For KdV equation (9), the relation between $n$ and $k$ of (13a) and (13b) is given as $k = n + 2$. Taking $n = 2$ and $k = 4$, leads to

$$\Phi(\eta) = a_0 + a_1 \varphi(\eta) + a_2 \varphi(\eta)^2, \tag{14a}$$

and

$$\frac{d\varphi}{d\eta} = \pm \sqrt{c_0 + c_1 \varphi(\eta) + c_2 \varphi(\eta)^2 + c_3 \varphi(\eta)^3 + c_4 \varphi(\eta)^4}. \tag{14b}$$

Using the symbolic software package Maple, the following solutions of the KdV equation (9), which describe the electrostatic potential, are given as:

$$\Phi(\eta) = a_0 + 3\frac{B}{A}\left(\frac{v - a_0 A}{B}\right) \operatorname{sech}\left(\frac{1}{2}\sqrt{\frac{v - a_0 A}{B}}\eta\right)^2, \tag{15}$$

$$\Phi(\eta) = a_0 + 3\frac{B}{A}\left(\frac{v - a_0 A}{B}\right) \sec\left(\frac{1}{2}\sqrt{\frac{a_0 A - v}{B}}\eta\right)^2, \tag{16}$$

$$\Phi(\eta) = a_0 + \frac{3}{2}\frac{B}{A}\left(\frac{v - a_0 A}{B}\right) \tanh\left(\frac{1}{2}\sqrt{\frac{a_0 A - v}{2B}}\eta\right)^2, \tag{17}$$

$$\Phi(\eta) = a_0 - \frac{3}{2}\frac{B}{A}\left(\frac{v - a_0 A}{B}\right) \tan\left(\frac{1}{2}\sqrt{\frac{v - a_0 A}{2B}}\eta\right)^2, \tag{18}$$

$$\Phi(\eta) = a_0 - 3\frac{B}{A}\left(\frac{v - a_0 A}{B}\right) \operatorname{csch}\left(\frac{1}{2}\sqrt{\frac{v - a_0 A}{B}}\eta\right)^2, \tag{19}$$

$$\Phi(\eta) = a_0 + 3\frac{B}{A}\left(\frac{v - a_0 A}{B}\right) \csc\left(\frac{1}{2}\sqrt{\frac{a_0 A - v}{B}}\eta\right)^2, \tag{20}$$

$$\Phi(\eta) = \frac{v}{A} - 12\frac{B}{A}\frac{1}{\eta^2}, \tag{21}$$

$$\Phi(\eta) = a_0 + 3\frac{m^2}{2m^2 - 1}\frac{B}{A}\left(\frac{v - a_0 A}{B}\right) cn\left(\frac{1}{2}\sqrt{\frac{a_0 A - v}{B(1 - 2m^2)}}\eta\right)^2, \tag{22}$$

$$\Phi(\eta) = a_0 - 3\frac{1}{m^2 - 2}\frac{B}{A}\left(\frac{v - a_0 A}{B}\right) dn\left(\frac{1}{2}\sqrt{\frac{a_0 A - v}{B(m^2 - 2)}}\eta\right)^2, \tag{23}$$

$$\Phi(\eta) = a_0 + 3\frac{m^2}{m^2 + 1}\frac{B}{A}\left(\frac{v - a_0 A}{B}\right) sn\left(\frac{1}{2}\sqrt{\frac{a_0 A - v}{B(m^2 + 1)}}\eta\right)^2, \tag{24}$$

$$\Phi(\eta) = a_0 - 3\frac{B}{A}(2c_3)^{2/3} \wp(C\eta; g_1; g_2), \tag{25a}$$

$$C = \sqrt[3]{c_3/4}, \qquad g_1 = -c_1/C \quad \text{and} \quad g_2 = -c_0, \tag{25b}$$

where $a_0$, $c_0$, $c_1$ and $c_3$ are arbitrary constants, $m$ is a modulus, $A$ and $B$ are given by (10) and $\wp(C\eta; g_1; g_2)$ is Weierstrass function of invariants $g_1$ and $g_2$.

## 7. Results and discussion

To make our result physically relevant, numerical studies have been made using plasma parameters close to those values corresponding to the dayside auroral zone [27].

However, since one of our motivations was to study effects of Mach number ($M$), the energetic population parameter ($\delta$) and the hot to cold electron equilibrium densities ratio ($\alpha$) on the existence of solitary waves by analyzing the Sagdeev's pseudo-potential for arbitrary amplitude electron-acoustic waves. Figures (1), (2) and (3) study the effects of M, $\delta$ and $\alpha$ parameters on the existence of solitary waves, respectively. In Figure (1), the behavior of $V(\phi)$ shows the critical Mach number for which a potential well in the negative-axis (corresponding to a solitary wave with a negative potential) develops with the going to upper limit of the Mach number for which a potential well in the negative-axis develops. Figure (2) shows that: changing the value of energetic population parameter varies the values of the critical and upper limit of Mach number.

For small amplitude electron-acoustic waves, the Korteweg-de Vries equation has been derived using the reductive perturbation method. A symbolic computational traveling wave method is used to obtain a series of exact solutions of the KdV equation. In Figures (4) and (5), a profile of the bell-shaped refractive solitary pulse is obtained for solutions (15) and (16). The soliton amplitude (width) increases (decreases) with the increase of the arbitrary constant $a_0$ as shown in Figure (5). Solutions (17) and (18) develop solitons with singularity at finite point called "blowup" of solutions. On other hand, solutions (19), (20) and the rational solution (21) lead to propagated explosive pulses as depicted in Figures (6), (7) and (8). Three Jacobi elliptic doubly periodic wave solutions are represented by the solutions (22), (23) and (24). These equations show that: when $m \to 1$, the Jacobi functions degenerate to the hyperbolic functions (15), (17) and (19), while when $m \to 0$ they degenerate to the triangular solutions (16), (18) and (20). Finally, solution (25) represents a Weierstrass elliptic doubly periodic solution.

To compare our result (the amplitude of the electrostatic potential) with that observed in the auroral zone, we choose a set of available parameters corresponding to the dayside auroral zone where an electric field amplitude $E_0 = 100\, mV/m$ has been observed [27] with $T_c \approx 5\, eV$, $T_h \approx 250\, eV$, $n_{c0} \approx 0.5\, m^{-3}$ and $n_{h0} \approx 2.5\, m^{-3}$. These parameters correspond to $\lambda_{Dh} \approx 7430\, cm$ and the normalized electrostatic wave potential amplitude $\phi_0 = \dfrac{E_0 \lambda_{Dh} e}{k_B T_h} \approx 0.03$, which is obtained for $\alpha = 5$, $v = 0.04$, $\delta = 0.48$ and $a_0 = 0.008$ [cf. Fig. (3)].

In summery, it has been found that the presence of non-thermal (fast) electrons modifies the properties of the electron acoustic solitary waves significantly and new exact solutions have been obtained. To our knowledge, these solutions have not been reported. It may be important to explain some physical phenomena in some plasma environments, such as terrestrial magnetosphere.

## 6. Conclusion

We have devoted quite some efforts to discuss the proper description of the new solutions and energetic population parameter ($\delta$) in unmagnetized collisionless plasma consisting of a cold electron fluid and non-thermal hot electrons obeying a non-thermal distribution, and stationary ions. The application of the pseudo-potential approach and the reductive perturbation theory to the basic set of fluid equations leads to Sagdeev's pseudo-potential form for arbitrary amplitude electron-acoustic waves and a KdV equation for small amplitude electron-acoustic waves. It is emphasized that amplitude of the electron-acoustic waves as well as parametric regime where the solitons can exist is sensitive to the Mach number ($M$), energetic population parameter ($\delta$) and the hot to cold electron equilibrium densities ratio ($\alpha$). Moreover, new solutions provide guidelines to classify the types of solutions according to the plasma parameters and can admit the following types of solutions: (a) polynomial solutions, (b) exponential solutions, (c) rational solutions, (d) triangular periodic wave solutions, (e) hyperbolic and solitary wave solutions. The arbitrary constant $a_0$ plays the role of consideration of higher-order approximation to increase (decrease) the amplitude and the width of the electron-acoustic wave solitons.

Solution (16) can be considered as a generalization of the solution has been obtained by El-Shewy [14]. In other word, the arbitrary constant $a_0$ play the same role of higher-order approximations in increasing (decreasing) the amplitude and width of (EAW) solitons. The rational solution (22) may be helpful to explain certain physical phenomena. Because a rational solution is a disjoint union of manifolds, particle systems describing the motion of a pole of rational solutions for a KdV equation were analyzed [28].

We have stressed out that it is necessary to include the magnetic field effect in the analysis to take the cold electron polarization effect into account to bring our model closer to observations. This is beyond the scope of the present paper and it will be include in a further work in electron-acoustic solitary wave. The application of our model might be particularly interesting in the auroral region.

## Figure Captions

Fig. (1): The behavior of Sagdeev's potential $V(\phi)$ vs $\phi$ for $\alpha=0.05$ and $\delta=0.03$ for different values of $M$.

Fig. (2): The behavior of Sagdeev's potential $V(\phi)$ vs $\phi$ for $\alpha=0.05$ and $M=1.2$ for different values of $\delta$.

Fig. (3): The behavior of Sagdeev's potential $V(\phi)$ vs $\phi$ for $\delta=0.03$ and $M=1.2$ for different values of $\alpha$.

Fig. (4): A bell-shaped solitary pulse represented by the solution (16) for different values of $a_0$ for $\alpha=5$, $\delta=0.1$, $v=0.04$.

Fig. (5): The variation of the amplitude and width of solitary pulse represented by the solution (17) for different values of $a_0$ for $\alpha=5$, $\delta=0.1$, $v=0.04$.

Fig. (6): An explosive pulse represented by the solution (20) for different values of $a_0$ for $\alpha=5$, $\delta=0.1$, $v=0.04$.

Fig. (7): An explosive pulse represented by the solution (21) for different values of $a_0$ for $\alpha=5$, $\delta=0.1$, $v=0.04$.

Fig. (8): An explosive pulse represented by the solution (22) for different values of $a_0$ for $\alpha=5$, $\delta=0.1$, $v=0.04$.

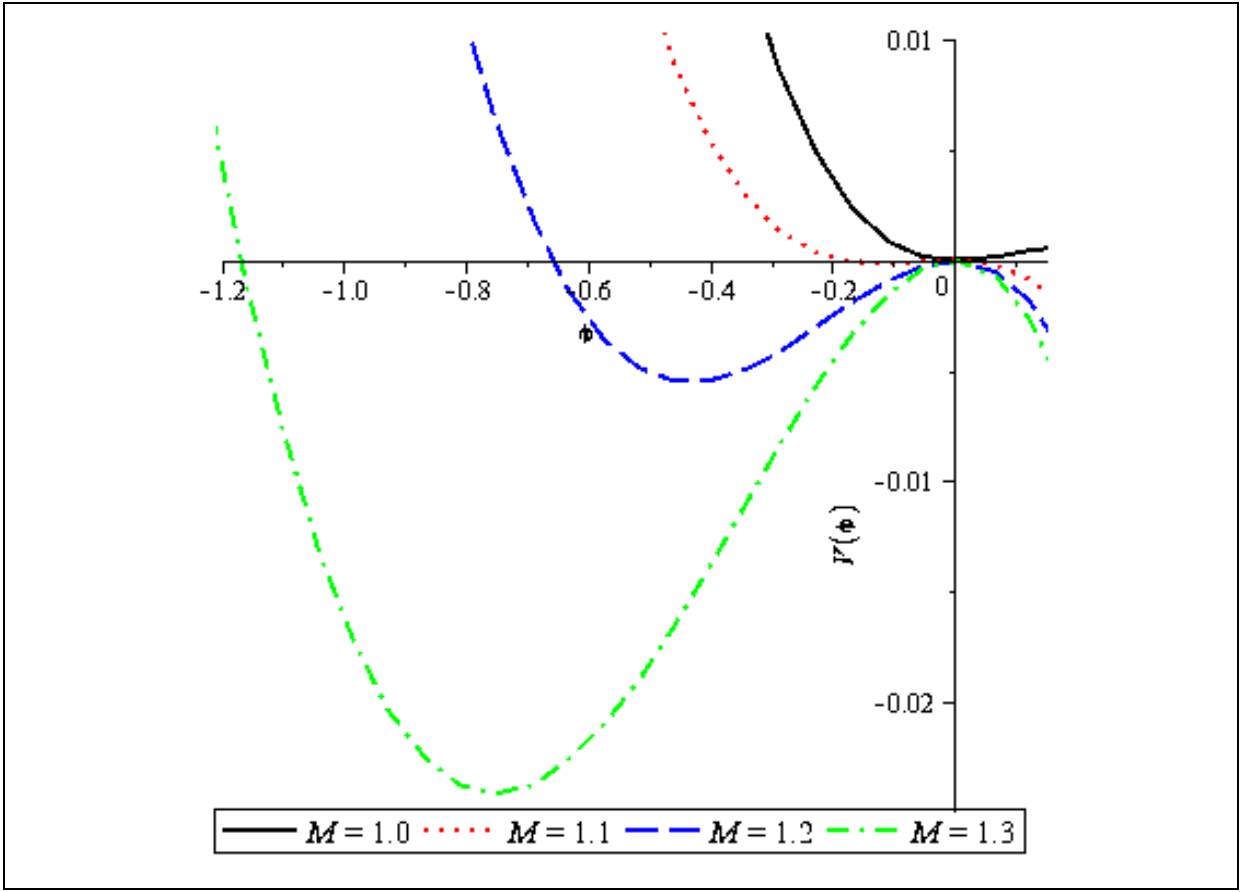

Fig. (1)

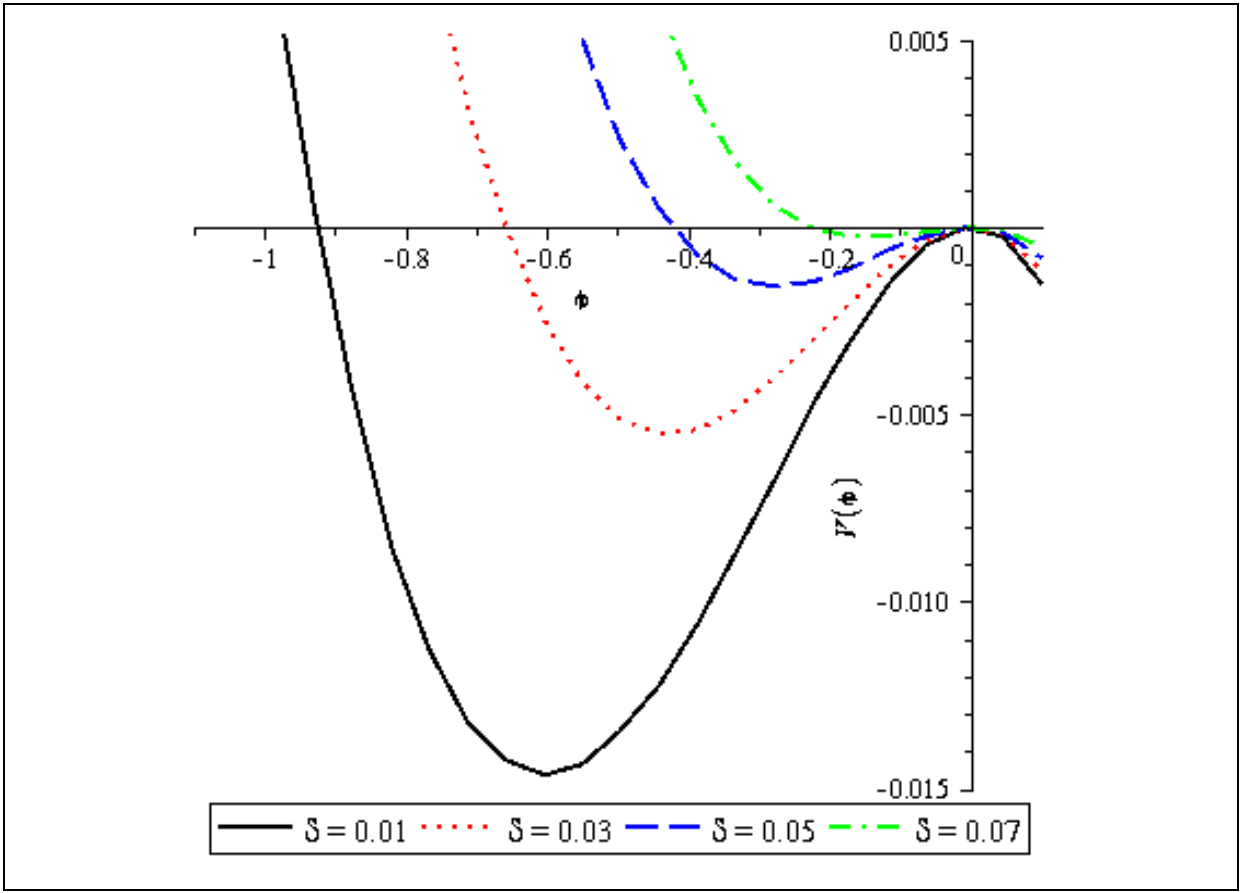

Fig. (2)

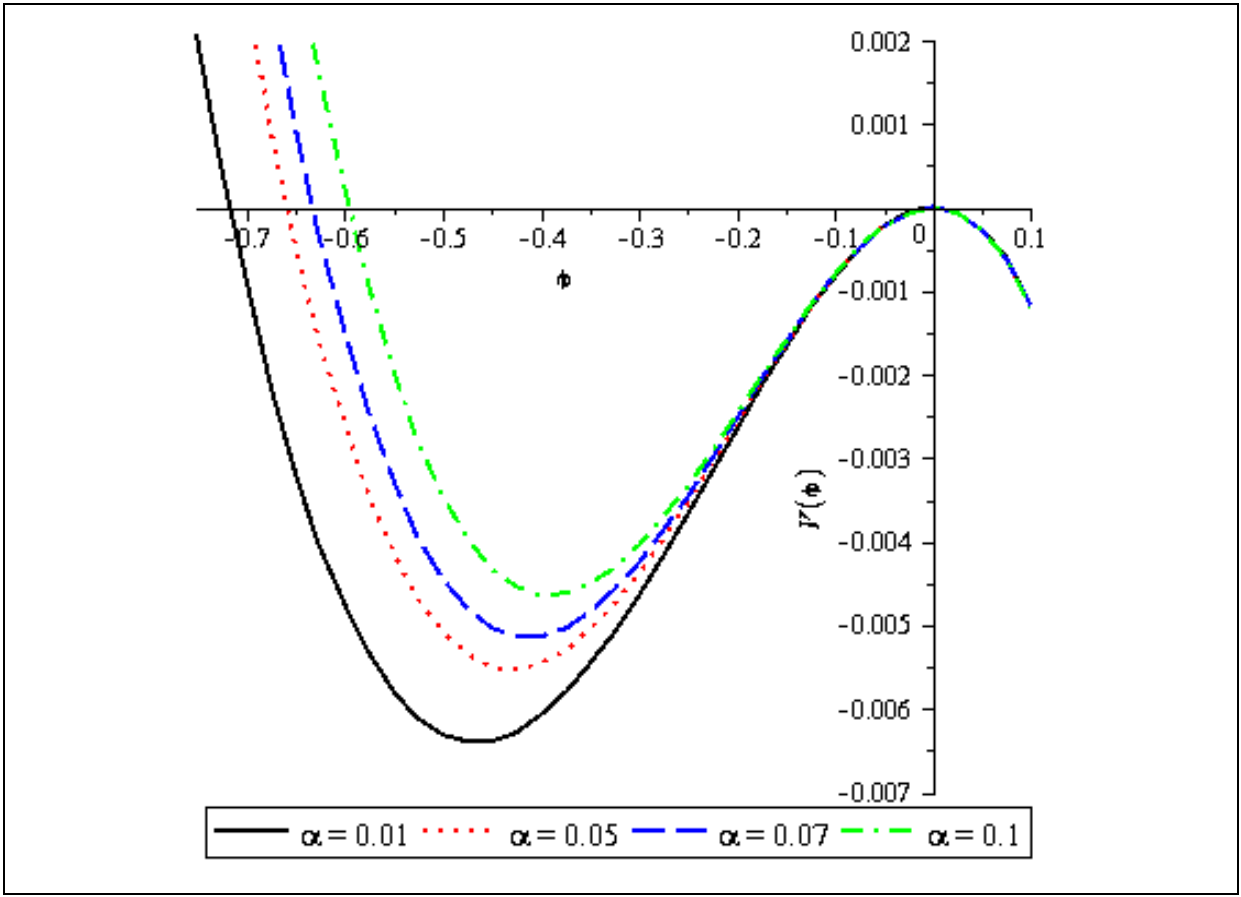

Fig. (3)

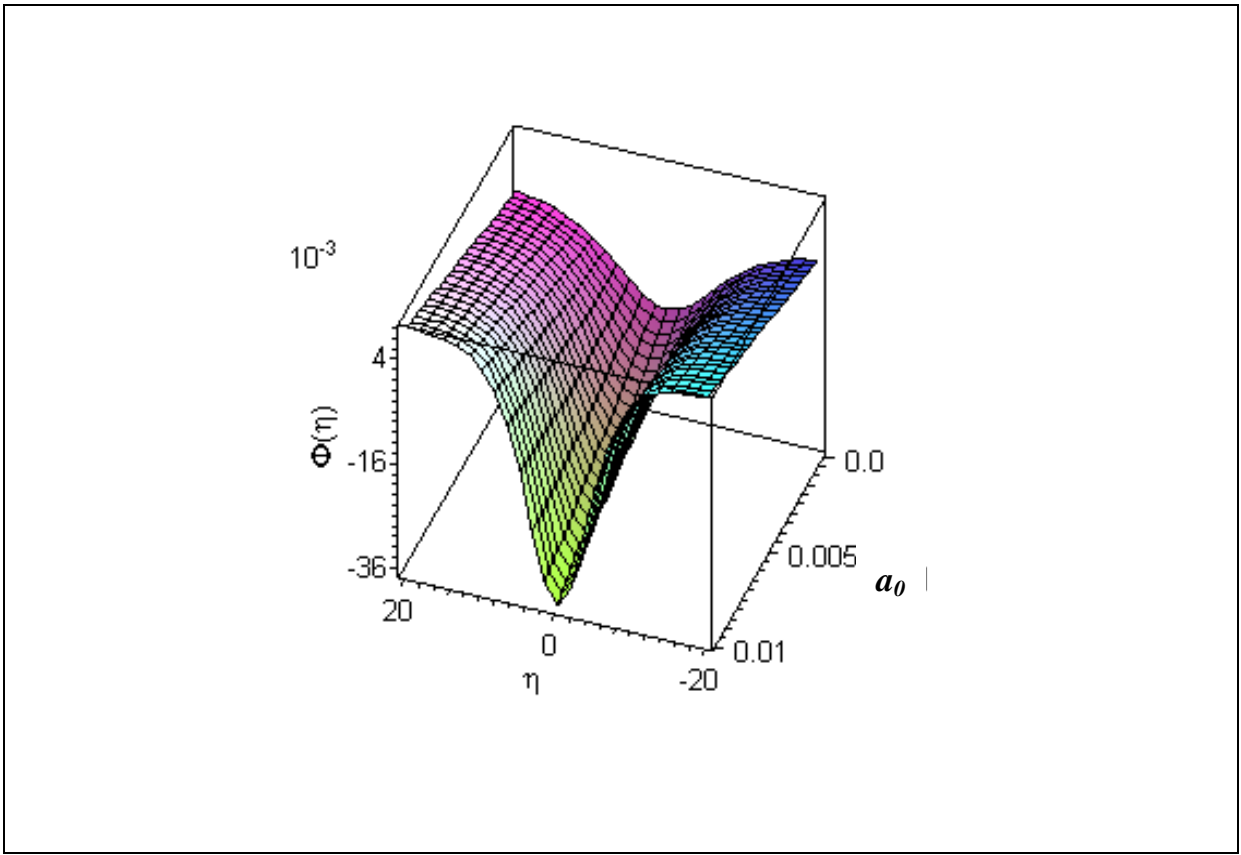

Fig. (4)

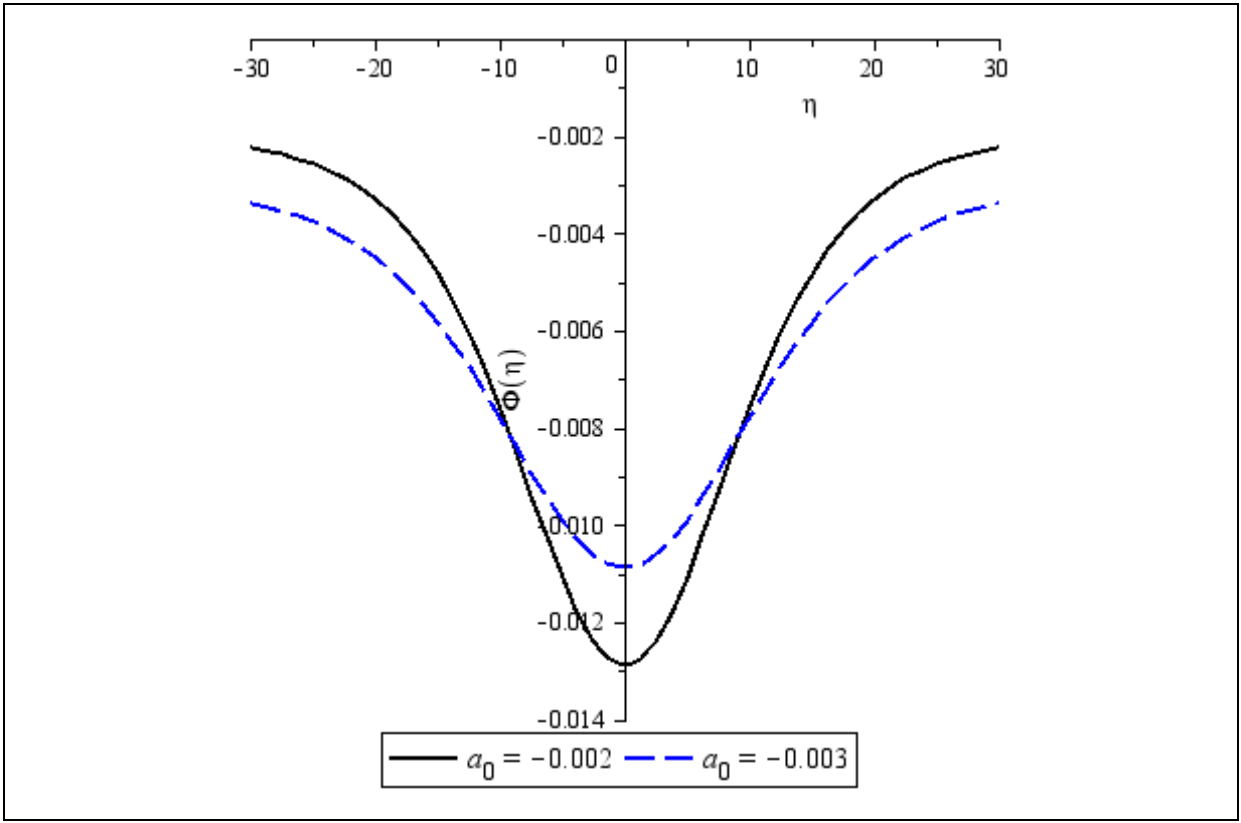

Fig. (5)

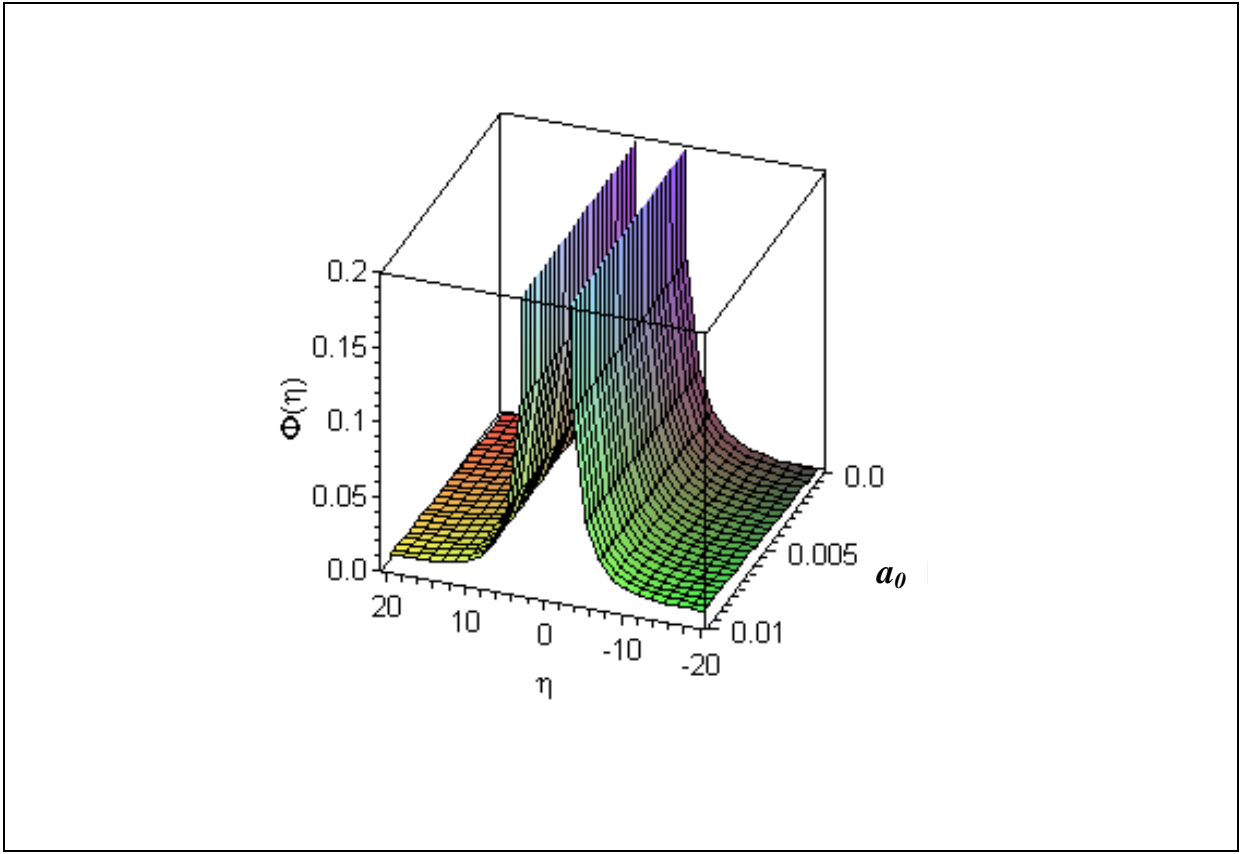

Fig. (6)

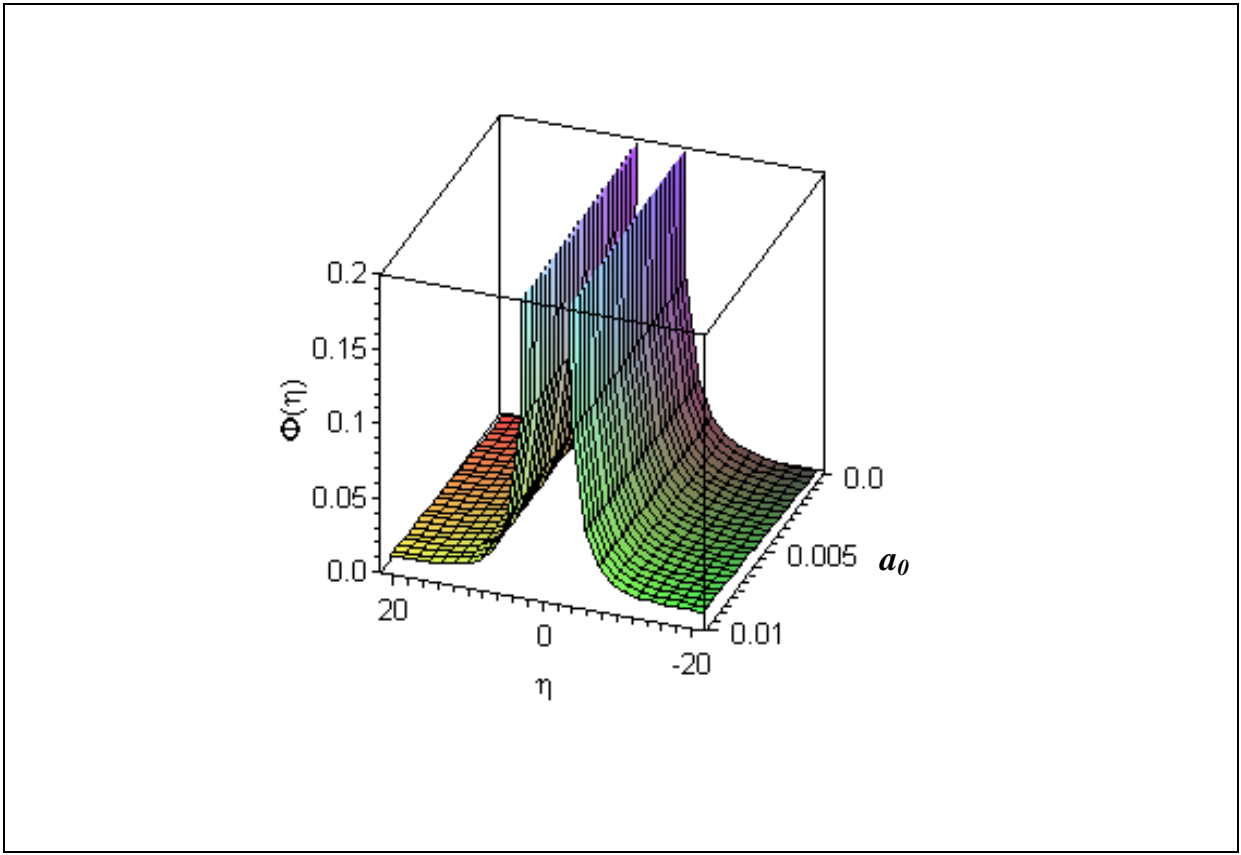

Fig. (7)

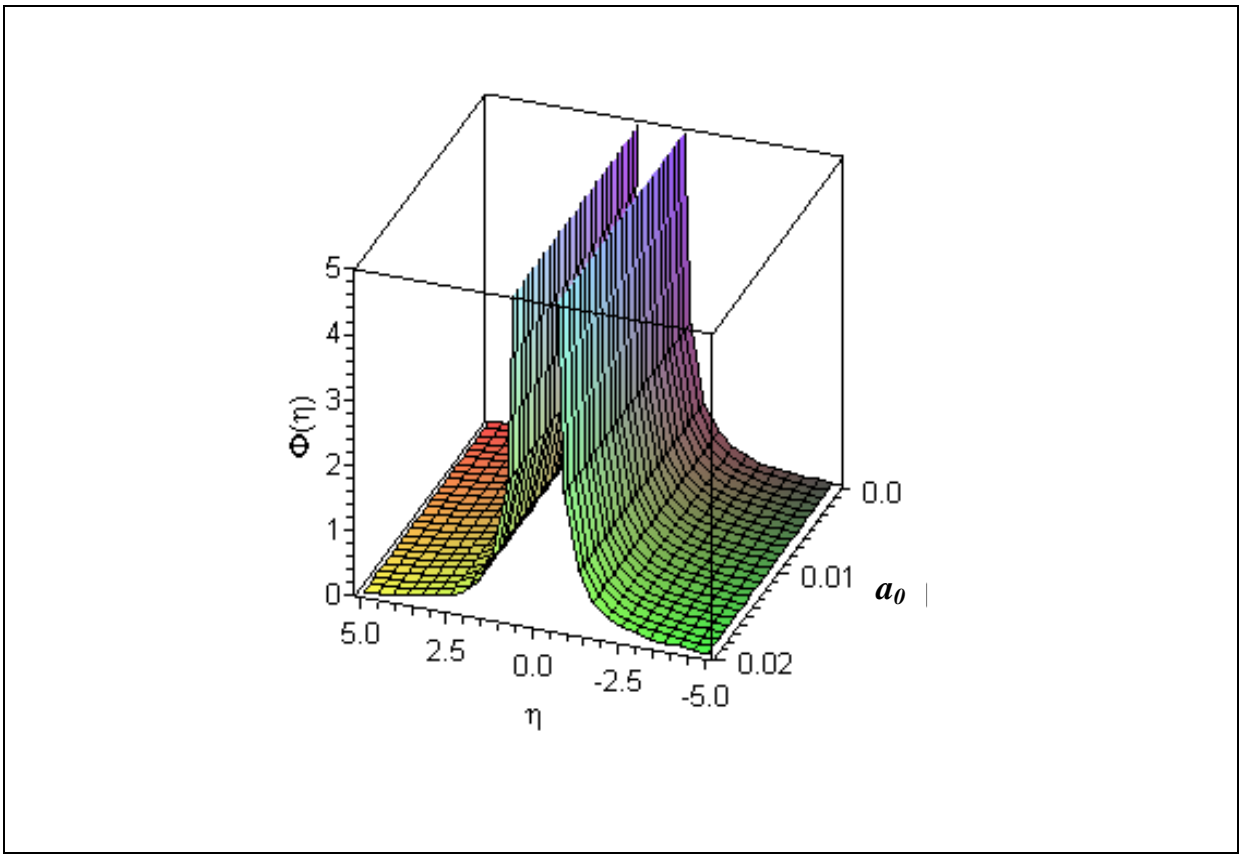

Fig. (8)